\documentclass[journal]{IEEEtran}
\usepackage{mathrsfs}
\usepackage{pifont}
\usepackage{bbding}
\usepackage{tipa}
\usepackage{amsmath,epsfig}
\usepackage{stmaryrd}
\usepackage{amssymb}
\usepackage{amsfonts}
\usepackage{epic}
\usepackage{graphicx}
\usepackage{curves}
\ifCLASSINFOpdf
\else
\fi
\begin{document}
\newtheorem{theorem}{Theorem}
\newtheorem{proposition}{Proposition}
\newtheorem{definition}{Definition}
\newtheorem{lemma}{Lemma}
\newtheorem{corollary}{Corollary}
\newtheorem{remark}{Remark}
\newtheorem{construction}{Construction}

\newcommand{\supp}{\mathop{\rm supp}}
\newcommand{\sinc}{\mathop{\rm sinc}}
\newcommand{\spann}{\mathop{\rm span}}
\newcommand{\essinf}{\mathop{\rm ess\,inf}}
\newcommand{\esssup}{\mathop{\rm ess\,sup}}
\newcommand{\Lip}{\rm Lip}
\newcommand{\sign}{\mathop{\rm sign}}
\newcommand{\osc}{\mathop{\rm osc}}
\newcommand{\R}{{\mathbb{R}}}
\newcommand{\Z}{{\mathbb{Z}}}
\newcommand{\C}{{\mathbb{C}}}
%

\title{A PAPR Reduction Method Based on Artificial Bee Colony Algorithm for OFDM Signals}
\author{{Yajun~Wang, Wen~Chen,~\IEEEmembership{Member,~IEEE} and
Chintha Tellambura,~\IEEEmembership{Senier Member,~IEEE}}
\thanks{Manuscript received January 12, 2010 and revised April 30, 2010. The associate editor coordinating
the review of this paper and approving it for publication was Murat
Uysal.}
\thanks{Yajun~Wang and Wen Chen are with Department of Electronic Engineering,
Shanghai Jiao Tong University, Shanghai, 200240 PRC. Yajun Wang is
also with the State Key Laboratory of Integrated Services Networks,
Xidian University; Wen Chen is also with SKL for Mobile
Communications, e-mail: \{wangyj1859;wenchen\}@sjtu.edu.cn.}
\thanks{Chintha Tellambura is with Department of Electrical and Computer
Engineering, University of Alberta, Edmonton, Canada,  T6G 2V4.
e-mail: chintha@ece.ualberta.ca.}
\thanks{This work is supported by NSF China \#60972031, by SEU SKL project
\#W200907, by ISN project \#ISN11-01, by Huawei Funding
\#YJCB2009024WL and \#YJCB2008048WL, and by National 973 project
\#2009CB824900.}}


%

\markboth{IEEE Transactions on Wireless Communications}{Shell
\MakeLowercase{\textit{et al.}}: Bare Demo of IEEEtran.cls for
Journals }
\maketitle

\begin{abstract}
One of the major drawbacks of orthogonal frequency division
multiplexing (OFDM) signals is the high peak to average power ratio
(PAPR) of the transmitted signal. Many PAPR reduction techniques
have been proposed in the literature, among which, partial transmit
sequence (PTS) technique has been taken considerable investigation.
However, PTS technique requires an exhaustive search over all
combinations of allowed phase factors, whose complexity increases
exponentially with the number of sub-blocks.
 In this
paper, a newly suboptimal method based on modified artificial bee colony
(ABC-PTS) algorithm is proposed to search the better combination of
phase factors. The ABC-PTS algorithm can significantly reduce the
computational complexity for larger PTS subblocks and offers lower
PAPR at the same time. Simulation results show that the ABC-PTS
algorithm is an efficient method to achieve significant PAPR
reduction .
\end{abstract}



%
\IEEEpeerreviewmaketitle

\section{Introduction}\label{sec:1}
In various high-speed wireless communication systems, the orthogonal
frequency division multiplexing (OFDM) has been used widely due to
its inherent robustness against multipath fading  and resistance to narrowband interference.
Well-known examples include wireless local area network (WLAN) IEEE 802.11a~\cite{IEEEconf:1} and wireless
metropolitan area network (WMAN) IEEE 802.16a~\cite{IEEEconf:2}, digital
audio broadcasting (DAB), digital
video broadcasting (DVB-T)~\cite{IEEEconf:3}.

However, one of the major drawbacks of OFDM signals is the high peak
to average power ratio (PAPR) of the transmitted signal. The high
peaks of an OFDM signal occur when the subsymbols for each
subcarrier are added up coherently. So OFDM signals can cause
serious problems including a severe power penalty
 at the transmitter which is particularly not affordable in portable wireless systems.
Several solutions have been proposed in recent years.
It is known that clipping~\cite{IEEEconf:4} is the simplest method,
but it degrades the bit-error-rate (BER) of the system, and results
in out-of-band noise and in-band distortion. Although
coding~\cite{IEEEconf:5,IEEEconf:6} can offer the best PAPR
reductions, the associated complexity and data rate reduction limit
the application of such a technique. On the other hand, selected
mapping (SLM) technique~\cite{IEEEconf:7} modifies the phases of the
original information symbols in each OFDM block and selects the
phase-modified OFDM block with the best PAPR performance for
transmission. However, the requirement of multiple IFFT operations
increases the implementation complexity.

In~\cite{IEEEconf:8,IEEEconf:9}, a tone reservation algorithm has
been proposed where several subcarriers are put apart  for PAPR
reduction. In~\cite{IEEEconf:9}, a tone injection algorithm has been
developed where the constellation points of part subcarriers are
modified to obtain PAPR reduction at the cost of an increase in
transmit power.
 An active set extension (ASE) algorithm has been proposed
in~\cite{IEEEconf:10,IEEEconf:11}. By modifying the exterior
modulation constellation over active subcarriers and not degrading
the BER performance, PAPR reduction is achieved.
In~\cite{IEEEconf:12}, a symmetric constellation extension (SCE)
algorithm has been developed for PAPR reduction, where  the
subsymbols for each subcarrier are represented by two symmetric
constellation points and an optimal representation has been derived
by using a derandomization algorithm. In~\cite{IEEEconf:13}, a
constellation extension method has been developed, where the data
for each subcarrier can be represented by a point in the original
constellation or by an extension point. By selecting an optimal
representation of the data points, PAPR reduction is obtained.
By modifying the modulation constellation or constellation
extension, these algorithms require an increase in the transmit
power and computation complexity at the transmitter.

The partial transmit sequence (PTS)~\cite{IEEEconf:14} is a
distortionless technique based on combining signal subblocks which
are phase-shifted by constant phase factors. The technique can get
sufficient  PAPR reduction and side information  need to be sent at
the same time. But the exhaustive search complexity of the optimal
phase combination increases exponentially with the number of
sub-blocks. So many suboptimal PTS methods have been developed. The
iterative flipping algorithm for PTS  in~\cite{IEEEconf:15}
has the computational complexity linearly proportional to the number
of subblocks. A neighborhood search is proposed
in~\cite{IEEEconf:16} using gradient descent search. A suboptimal
method in~\cite{IEEEconf:17} is developed by modifying the problem
into an equivalent problem of minimizing the sum of phase-rotated
vectors. A simulated annealing method is proposed in~\cite{IEEEconf:19}.
A suboptimal PTS algorithm based on particle swarm optimization  is proposed
in~\cite{IEEEconf:20,IEEEconf:21}. An intelligent genetic algorithm for PAPR
reduction is developed in~\cite{IEEEconf:22,IEEEconf:23}.

In this paper, we propose a newly suboptimal phase optimization scheme based on
modified artificial bee colony (ABC-PTS) algorithm, which can
efficiently reduce the PAPR of OFDM signals. The proposed scheme can
search the better combination of the initial phase factors.
Simulation results show that the ABC-PTS phase optimization scheme
can achieve superior PAPR reduction performance and at the same time
requires far less  computational complexity than the previous PTS
techniques. Like the original PTS, our scheme  also requires to send
side information.

This paper is organized as follows. In Section II, definition of
PAPR of OFDM signals and the complementary cumulative distribution function (CCDF)
 are introduced. The principles of PTS techniques are described in Section III. The modified ABC (ABC-PTS)
algorithm to search the better combination of  the phase factors is
proposed in Section IV. In Section V, the performance of ABC-PTS
algorithm and other algorithms for PAPR reduction is evaluated by
computer simulation. Conclusions are made in Section VI.

\section{OFDM System And PAPR}\label{sec:2}
In an OFDM system, a high-rate data stream is split into $N$ low-rate streams that are
 transmitted simultaneously by subcarriers, where
$N$ is  the number of subcarriers. Each of the subcarriers is independently modulated using
  phase-shift keying (PSK) or quadrature amplitude modulation (QAM).
The  inverse discrete Fourier transform (IDFT)  generates
  the ready-to-transmit  OFDM signal.
For an input OFDM block $\textbf{X}=[X_0,\dots,X_{N-1}]^T$,  each symbol in $\textbf{X}$
modulates one
 subcarrier of $\{f_0,\dots,f_{N-1}\}$. The $N$ subcarriers are orthogonal,
 i.e, $f_n=n\Delta f$, where $\Delta f =
 1/NT$ and $T$ is the symbol period. The complex envelope of the transmitted OFDM signal in one symbol period
 is given by
\begin{equation}\label{eq1}
x(t)=\frac{1}{\sqrt{N}}\sum_{n=0}^{N-1}X_n e^{j2\pi {f_n}t},0\leq t<
{NT}.
\end{equation}

The PAPR of $x(t)$ is defined as the ratio of the maximum
instantaneous power to the average power, that is
\begin{equation}\label{eq2}
PAPR=\frac{\underset{0\leq t<NT}{\max}|x(t)|^2}{E[|x(t)|^2]},
\end{equation}
where
\begin{equation}\label{eq3}
E[|x(t)|^2]=1/{NT}\int_{0}^{NT}{|x(t)|^2} dt.
\end{equation}
However, most systems use discrete-time signals in which the OFDM
signal is expressed as
\begin{equation}\label{eq4}
x(k)=\frac{1}{\sqrt{N}}\sum_{n=0}^{N-1}X_n\cdot e^{\frac{j2\pi
nk}{LN}},k=0,1,\cdots,LN-1,
\end{equation}
where $L$ is the oversampled factor. It has been shown
in~\cite{IEEEconf:18} that the oversampled factor $L=4$ is enough
to provide a sufficiently accurate estimate of the PAPR of OFDM
signals.

The complementary cumulative distribution function (CCDF) is one of
the most frequently used performance measures for PAPR reduction,
representing the probability that the PAPR of an OFDM symbol exceeds
the given threshold $PAPR_{0}$, which is denoted as
\begin{equation}\label{eq5}
 CCDF=Pr(PAPR>PAPR_{0}).
\end{equation}
\section{PTS TECHNIQUES}\label{sec:3}
\begin{figure}
\centering
\includegraphics[width=3.5in,angle=0]{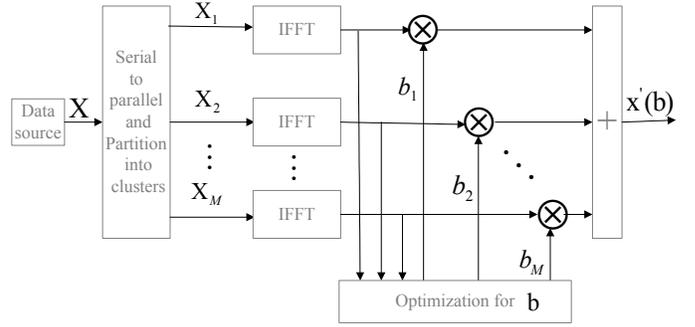}
\caption{Block diagram of the PTS technique.} \label{fig1}
\end{figure}

The principle structure of PTS method is shown in Fig.~\ref{fig1}.
     The input data block $\textbf{X}$ is partitioned into $M$ disjoint
    sub-blocks $\textbf{X}_m,m=1,2,\dots M$ such that $\textbf{X}=\sum\limits_{m=1}^M\textbf{X}_m$.
    Sub-blocks are combined to minimize the PAPR in the time domain. $L$-times oversampled time domain
    signal of $\textbf{X}_m$ is
denoted as $\textbf{x}_m,m=1,2,\dots M$, which are obtained by
taking an IDFT of length $NL$ on $\textbf{X}_m$ concatenated with
$(L-1)N$ zeros. Each $\textbf{x}_m$ is multiplied by a phase
weighting factor $b_m=e^{j\phi_m}$, where~$\phi_m\in [0,2\pi)$~for
~$m=1,2,\dots M $. The goal of the PTS approach is to find an
optimal phase weighted combination to minimize the PAPR value. The
transmitted signal in the time domain after combination can be
expressed as
\begin{equation}\label{eq6}
\textbf{x}^{'}(\textbf{b})=\sum_{i=1}^{M}b_i\textbf{x}_i,
\end{equation}
where
$\textbf{x}^{'}(\textbf{b})=[x_{1}^{'}(\textbf{b}),x_{2}^{'}(\textbf{b}),\cdots,x_{NL}^{'}(\textbf{b})]$.

In general, the selection of the phase factor is limited to a set
with finite number of elements to reduce the search complexity. The
set of allowed phase factors is
\begin{equation}\label{eq7}
\textbf{P}=\{e^{j2\pi\ell/W}|\ell=0,1,\dots,W-1\}.
\end{equation}
where $W$ is the number of allowed phase factors. We can fix a phase
factor without any performance loss. There are only $M-1$ free
variables to be optimized and hence $W^{M-1}$ different phase
vectors are searched to find the global optimal  phase factor. The
search complexity increases exponentially with $M$, the number of
sub-blocks.

\section{Minimize PAPR Using Modified ABC algorithm}
In order to get the OFDM signals with the minimum PAPR, a suboptimal
combination method based on the modified artificial bee colony (ABC) algorithm is proposed to solve the optimization problem of PTS. The modified ABC algorithm
with lower complexity can get better PAPR performance.

The minimum PAPR for PTS method is relative to the problem:

\noindent Minimize
\begin{equation}\label{eq8}
f(\textbf{b})=
\frac{\max|x^{'}(\textbf{b})|^2]|}{E[|x^{'}(\textbf{b})|^2]},
\end{equation}
subject to 
\begin{equation}\label{eq9}
\textbf{b}\in\{e^{j\phi_m}\}^{M},
\end{equation}
where $\phi_m\in \{\frac{2\pi k}{W}|k=0,1,\dots,{W-1}\}$.

\subsection{ Artificial Bee Colony Algorithm}
In recent years, Karaboga $et~al$
\cite{IEEEconf:24,IEEEconf:25,IEEEconf:26} introduced a bee swarm
algorithm called artificial bee colony (ABC) algorithm for numerical
optimization problems.  In the ABC algorithm, the colony of
artificial bees contains three groups of bees: employed bees,
onlookers and scouts. Each cycle of the search consists of three
steps: (1) placing the employed bees onto the food sources and then
calculating their nectar amounts; (2) selecting the food sources by
the onlookers after sharing the information of employed bees and
determining the nectar amount of the foods; (3) determining the
scout bees and placing them onto the randomly determined food
sources. In the ABC, a food source position represents a possible
solution to the problem to be optimized and the nectar amount of a
food source corresponds to the quality (fitness) of the associated
solution.


At the initialization step, a set of food source positions are
randomly produced and corresponding nectar amounts are calculated.
Each employed bee is moved onto her food source area for determining
a new food source  within the neighbourhood of the present one, and
then its nectar amount is evaluated. If the nectar amount of the new
one is higher than that of the previous one, she memorizes the new
position and forgets the old one. Otherwise she keeps the position
of the previous one. After all employed bees complete the search
process, they come back into the hive and share the nectar
information of the food sources (solutions) and their position
information with the onlooker bees waiting on the dance area. All
onlookers determine a food source area with a probability based on
their nectar amounts. If the nectar amount of a food source is much
higher when compared with other food sources, this means that this
source will be chosen by most of the onlookers. Each onlooker
determines a neighbourhood food source within the neighbourhood of
the one to which she has been assigned and then its nectar amount is
evaluated. The selection of the scout bee is controlled by a control
parameter called \textquotedblleft limit\textquotedblright. If a
solution representing a food source cannot be improved by a
predetermined number of trials, i.e., \textquotedblleft
limit\textquotedblright, it means that the associated food source
has been exhausted by the bees and then the employed bee of this
food source becomes a scout. The position of the abandoned food
source is replaced with a randomly produced food position. So
\textquotedblleft limit\textquotedblright~ controls the selection of
the scout bee and the qualities of solutions. These three steps are
repeated until the termination criteria are satisfied. For a
complete understanding of the ABC method, the reader is referred
to~\cite{IEEEconf:24,IEEEconf:25,IEEEconf:26}.

\subsection{Modified Artificial Bee Colony Algorithm to Reduce PAPR}

  Due to the facts that the original ABC algorithm is only suitable for continuously numerical optimization problems,
 we have to  do some modifications for the original ABC algorithm in order to apply ABC algorithm
 to search the better combination of phase factors for PTS. We refer to the modified ABC algorithm as
 ABC-PTS. In the paper, we select the phase factor $\textbf{b}=\{-1,1\}^M$
 or $\textbf{b}=\{-1,1,j,-j\}^M$.

 In the ABC-PTS algorithm, a
food source position represents a phase vector
$\textbf{b}_i=[b_{i1},b_{i2},\cdots,b_{iM}]^T,~i=1,2,\cdots,S $,
~where
  $S$ denotes the size of a randomly distributed initial population.
  The nectar amount of a food source or fitness value of a solution $\textbf{b}_i$ in the population is
   determined by the following formula:
 \begin{equation}\label{eq10}
fitness(\textbf{b}_i)=\frac{1}{1+f(\textbf{b}_i)}.
\end{equation}

  For each employed bee, a candidate food source position from the
previous one is produced by the following formula:
\begin{equation}\label{eq11}
b^{'}_{il}=b_{il}+\phi_{il}(b_{il}-b_{kl}),
\end{equation}
where $l\in\{1,2,\cdots,M\}$ and $k\in\{1,2,\cdots,J\}$, $i\neq k$,
$J$ is the number of employed bees (the number of food sources), 
and $\phi_{il}$ is a random number between [-1,1]. Due to $b_{il}^{'}$ is the discrete
coordinate, thus (\ref{eq11}) is modified to the following formulas:

For $W=2$
\begin{equation}\label{eq12}
b^{'}_{il}=\left\{\begin{array}{lcr}1, & if~~ \pi/4\leq
b^{'}_{il}<5\pi/4,\\ -1, & ~~ else,\end{array}\right.
\end{equation}

For $W=4$
\begin{equation}\label{eq13}
b^{'}_{il}=\left\{\begin{array}{lcr}j, & if~~ \pi/4\leq
b^{'}_{il}<3\pi/4,\\ -1, & if ~~ 3\pi/4\leq b^{'}_{il}<5\pi/4,\\ -j,
& if ~~ 5\pi/4\leq
b^{'}_{il}<7\pi/4,\\
1,& ~~ else,\end{array}\right.
\end{equation}

  For each onlooker bee,  a food source is chosen depending on the probability
value associated with that food source, $p_i$, calculated by the
following formula:
\begin{equation}\label{eq14}
p_{i}=\frac{fitness(\textbf{b}_i)}{\sum\limits_{i=1}^{S}
fitness(\textbf{b}_i)}.
\end{equation}

  After all onlookers are distributed onto the food sources and their nectars are tested, sources are
checked whether they are to be abandoned. If the number of cycles
that a source can not be improved is greater than a predetermined
limit, the source is considered to be exhausted. The employed bee
associated with the exhausted source becomes a scout and makes a
random search in problem domain by the following formula:
\begin{equation}\label{eq15}
b_{il}=b^{min}_{l}+(b^{max}_{l}-b^{min}_{l})*rand,
\end{equation}

Our proposed modified ABC algorithm for PAPR reduction (ABC-PTS) can
thus be summarized as follows.

\begin{enumerate}\it
\item  Initialize food source positions, set the value of limit and the maximum iteration number.


\item  Determine  neighbour food source positions for the employed bees using (\ref{eq11}). Then modify
food source positions using (\ref{eq12}) or (\ref{eq13}).

\item  Calculate the nectar amounts or fitness using (\ref{eq10}).

\item  If all onlookers are assigned food sources, go to Step~7.
Otherwise, continue.

\item  Select a food source for an onlooker using (\ref{eq14}).

\item  Determine a neighbour food source position for the onlooker using (\ref{eq11}). Then modify
food source positions using (\ref{eq12}) or (\ref{eq13}). Go to Step
4.

\item  Find the abandoned food source and allocate its employed bee as scout for searching new food
sources using (\ref{eq15})
\item  Memorize the position of the best food source.

\item  If the maximum iteration number is reached, output final food
source positions and stop. Otherwise go to Step 2.
\end{enumerate}

\subsection{Complexity Analysis for ABC-PTS and the Existing PAPR Reduction Methods }

In~\cite{IEEEconf:15}, the iterative flipping algorithm for PTS
(IPTS) was proposed for PAPR reduction. The method has the
computational complexity linearly proportional to the number of
subblocks, i.e. the search complexity is proportional to $(M-1)W$. A
neighborhood search using gradient descent search (GD) is proposed in~\cite{IEEEconf:16}.
The technique first sets the initial phase factor
$\textbf{b}=[1,1,\cdots,1]$ and the number of maximum iteration $I$,
then searches the phase factor that achieves the smallest PAPR in
the neighbour of $\textbf{b}$ with radius $r$.  The search
complexity of this method is proportional to $C^{r}_{M-1}W^{r}I$,
where $C^{m}_{n}$ is the binomial coefficient. A suboptimal method (TS)
in~\cite{IEEEconf:17}  is developed by modifying the problem into an
equivalent problem of minimizing the sum of phase-rotated vectors.
The phase factor of the method is continuously changed in $[0,2\pi]$.
The search complexity of this method is proportional to $LN$, where
$N$ is the number of subcarrier and $L$ is the oversampled factor.
In~\cite{IEEEconf:20,IEEEconf:21}, a particle swarm optimization algorithm (PSO-PTS) is
proposed to reduce PAPR. The search complexity of this method is proportional to $SG$, where
 $S$ is the size of particle swarm, $G$ is the maximal generations of PSO-PTS.
 An intelligent genetic algorithm (GA) called minimum distance guided GA (MDGA) is
developed in~\cite{IEEEconf:22,IEEEconf:23}. The MDGA generates initial population by using the
output of the IPTS, perturbing the output of the IPTS with minimum Hamming Distance and mutating the
 output of the IPTS randomly. Then MDGA search the phase factor by an intelligent  replacement  strategy,
 crossover and mutation.  The search complexity of this method is proportional to $PG+(M-1)W$, where $P$ is
 the size of the population, $G$ is the maximal generations of MDGA.
In the ABC-PTS algorithm,  the randomly initial phase factor
population with the size $S$ are produced, then all employed bees
and onlookers carry out search according to the algorithm, when the
maximum iteration number $K$ is reached, the phase factor with the
minimum PAPR is thought as the approximately optimal one. So the
search complexity of this method is proportional to $SK$. The
complexity of the PTS technique with an exhaustive search (OPTS)~\cite{IEEEconf:14}
 is $W^{M-1}$ by fixing a phase factor without any performance loss.
\begin{figure}
\centering
\includegraphics[width=3.5in,angle=0]{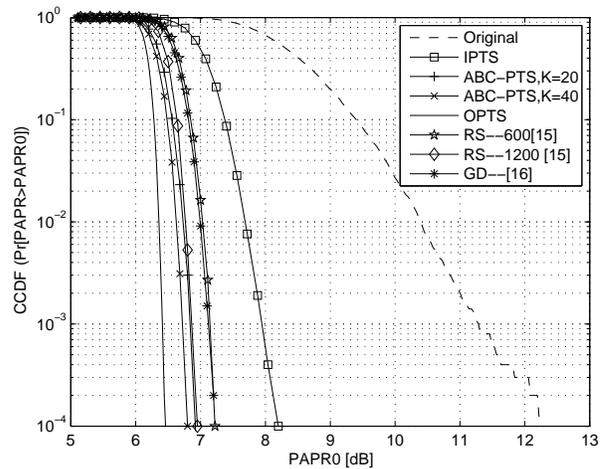}
\caption{Comparison of PAPR reduction among ABC-PTS with different iterations and the other methods, W=2.}
\label{fig2}
\end{figure}

\section{Simulation results}
To evaluate and compare the performance of the ABC-PTS algorithm for
OFDM PAPR reduction, numerous simulations have been conducted.
 In order to get CCDF, 100000 random OFDM symbols are generated. The
transmitted signal is oversampled by a factor of $L=4$ for accurate
PAPR. In our simulation, 16-QAM modulation with $N=256$ sub-carriers
is used and the phase factor $W=2$ is chosen. When larger phase
factor, for example, $W=4$ is chosen, the similar simulation results
can be obtained, while the performance will be better.


In the ABC-PTS algorithm, there are three control parameters: the
number of the food sources, the value of  limit and
 the maximum iteration number. Employed bees or onlooker bees carry out the exploiting
process in the search space, the scouts control the exploration
process in the ABC-PTS algorithm. The two processes are implemented
together. Different maximum iteration number, different size of
population and different limit value are chosen to evaluate the
performance of the ABC-PTS algorithm for PAPR reduction. In
simulation, $S$ denotes the number of the food sources or the size
of population, $K$ denotes the maximum iteration number, $Limit$
denotes the value of limit.

In Fig.~\ref{fig2},  the CCDF for $M=16$ sub-blocks using random partition is
shown. Here $S=30$, $Limit=5$ and  different iteration $K=20$,
$K=40$ for the ABC-PTS. When $P_r(PAPR>PAPR_0)=10^{-3}$, the PAPR of
the original OFDM is $11.3$\,dB. The PAPR by IPTS is $7.95$\,dB.
The PAPR by the ABC-PTSs with iteration number $20$ and $40$ are
approximately $6.75$\,dB and $6.65$\,dB respectively. Using the random search
(RS) in~\cite{IEEEconf:15}, when the numbers of randomly selected phase factors are $600$ and $1200$,
the PAPRs are reduced to $7.15$\,dB and $6.8$\,dB respectively. The PAPR by the
gradient descent search (GD)  with the search complexity
$C^{r}_{M-1}W^{r}I=C^{2}_{15}2^{2}3=1260$ in~\cite{IEEEconf:16} is
$7.1$\,dB.  The PAPR by the OPTS with exhaustive search number $2^{15}=32768$ is
$6.4$\,dB. There is a $0.25$\,dB gap between the PAPR by OPTS and
by ABC-PTS with iteration number $K=40$. But from the analysis in
section IV-C, we can know that the search complexity of the ABC-PTS
with $K=40$ is only $SK/W^{(M-1)}=1200/32768=3.66\%$ of that by the
OPTS. For the same or almost same search complexity, the performance
of the ABC-PTS with $K=40$ is also better than that of RS and GD.

Table.~\ref{Table1} shows comparison of computational complexity among
different methods for $M=16$ subblocks, where the size of population for
PSO-PTS~\cite{IEEEconf:20,IEEEconf:21}, MDGA~\cite{IEEEconf:22,IEEEconf:23} and ABC-PTS are $S=P=30$,
the number of maximal generations or iterations are $G=K=30$.
It can be seen that the performance of  swarm intelligence algorithms, i.e.PSO-PTS, MDGA and ABC-PTS
excels that of  other methods. For the  same search
complexity, the PAPR of the ABC-PTS  is smaller $0.3$\,dB than that
 of PSO-PTS. For the almost  same search
complexity, the PAPR of the ABC-PTS  is smaller $0.2$\,dB than that
 of MDGA.
\begin{table}[h]
\begin{center}
\caption {When $CCDF=10^{-3}$, COMPARISON OF COMPUTATIONAL COMPLEXITY  AMONG DIFFERENT METHODS FOR PHASE FACTOR $W=2$, $M=16$ SUB-BLOCKS,
SIZE OF POPULATION/PARTICLE $P=S=30$ AND MAXIMAL GENERATIONS/ITERATIONS $G=K=30$}\label{Table1}
\begin{tabular}{|c|c|c|}

\hline    methods & computational complexity & PAPR\\
\hline     IPTS     &$(M-1)W=15*2=30$                            &7.95\,dB \\
\hline     GD       &$C^{r}_{M-1}W^{r}I=C^{2}_{15}2^{2}3=1260$   &7.15\,dB  \\
 \hline    TS       &$LN=4*256=1024$                             &7.25\,dB \\
\hline     PSO-PTS  &$SG=30*30=900$                              &7.1\,dB \\
\hline     MDGA   &$PG+(M-1)W=30*30+15*2=930$                    &7.0\,dB \\
\hline     ABC-PTS  &$SK=30*30=900$                              &6.8\,dB \\
\hline     OPTS     &$W^{M-1}=2^{15}=32768$                      &6.45\,dB \\
 \hline
\end{tabular}
\end{center}
\end{table}

\begin{figure}
\centering
\includegraphics[width=3.5in,angle=0]{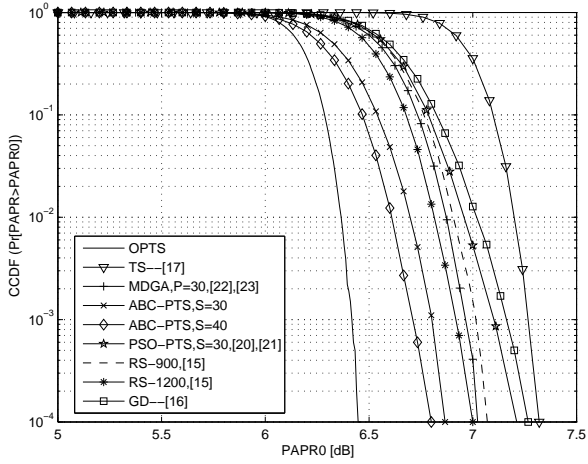}
\caption{Comparison of PAPR reduction among ABC-PTS with different size of
population and the other methods, W=2.} \label{fig3}
\end{figure}

In Fig.~\ref{fig3}, we compare the PAPR reduction performance of the
ABC-PTS with the  other methods
in~\cite{IEEEconf:15,IEEEconf:16,IEEEconf:17,IEEEconf:20,IEEEconf:21,
IEEEconf:22,IEEEconf:23} for the same or almost
same search complexity.
Fig.~\ref{fig3} shows the simulation results of the ABC-PTS
 with different size of population, the same maximum iteration number $K=30$ and $Limit=5$, where subblocks
 $M=16$ are generated by random partition.
 When $P_r(PAPR>PAPR_0)=10^{-3}$, the PAPR by OPTS is approximately $6.45$\,dB.
 By using the ABC-PTS with  $S=30$ and $S=40$, the
 PAPRs are reduced to  $6.8$\,dB and $6.7$\,dB, respectively.
Compared to the PAPR by OPTS, the PAPR by the ABC-PTS with
$S=30$ and $S=40$ has a gap approximately  $0.35$\,dB and
$0.25$\,dB, respectively. But  the search complexity of the ABC-PTS
 is only  $2.75\%$ and $3.66\%$ of that by the OPTS, respectively.  Using RS
 in~\cite{IEEEconf:15}, when the numbers of randomly selected phase factors are $900$ and $1200$, the PAPRs are reduced to  $7$\,dB and $6.9$\,dB,
respectively. The PAPR by GD with the search complexity
$C^{r}_{M-1}W^{r}I=C^{2}_{15}2^{2}3=1260$ in~\cite{IEEEconf:16} is $7.15$\,dB. The PAPR by TS with the search complexity $LN=4*256=1024$
in~\cite{IEEEconf:17} is $7.3$\,dB.   Using the MDGA  with the search complexity
$PG+(M-1)W=30*30+15*2=930$ in~\cite{IEEEconf:22,IEEEconf:23},
the PAPR is reduced to $6.95$\,dB. The PAPR by the PSO-PTS  with the search complexity $SG=30*30=900$ in~\cite{IEEEconf:20,IEEEconf:21} is
$7.1$\,dB.  From Fig.~\ref{fig3}, it can  be seen that apart from the PAPR by OPTS , the PAPR
reduction performance of the ABC-PTS is the best among that of all
methods for the same or almost same search complexity.

\begin{figure} \centering
\includegraphics[width=3.5in,angle=0]{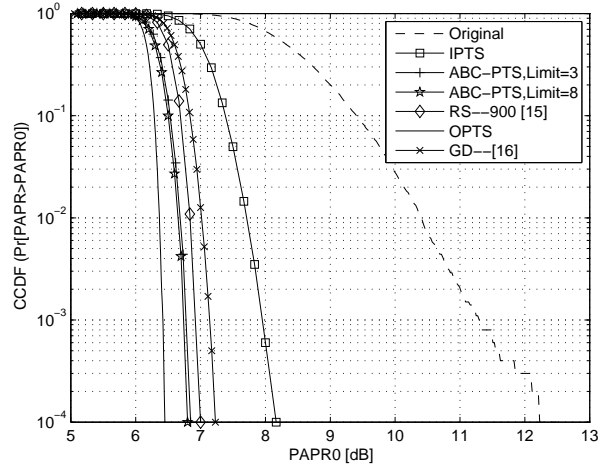}
\caption{Comparison of PAPR reduction among ABC-PTS with different Limit and the other methods, W=2.}
\label{fig4}
\end{figure}
In Fig.~\ref{fig4}, we compare the PAPR reduction performance of the
ABC-PTS with different Limit, the same size of population $S=30$ and
the same maximum iteration number $K=30$ for $M=16$ sub-blocks .
When $P_r(PAPR>PAPR_0)=10^{-3}$, the PAPR of the original OFDM is
$11.3$\,dB, the PAPRs by the ABC-PTS with $Limit =3$ and $Limit =8$
are $6.8$\,dB and $6.8$\,dB respectively. The PAPR by the OPTS is $6.5$ dB. The
PAPR by IPTS is $7.95$\,dB. The PAPR by RS~\cite{IEEEconf:15} with 900 randomly
selected phase factors is $6.95$\,dB. The PAPR by GD~\cite{IEEEconf:16} is $7.1$\,dB. From Fig.~\ref{fig4}, it can be discovered that the difference of
the PAPR between $Limit =3$ and $Limit =8$ is negligible. Little
performance improvement can be obtained by increasing Limit.

\begin{figure} \centering
\includegraphics[width=3.5in,angle=0]{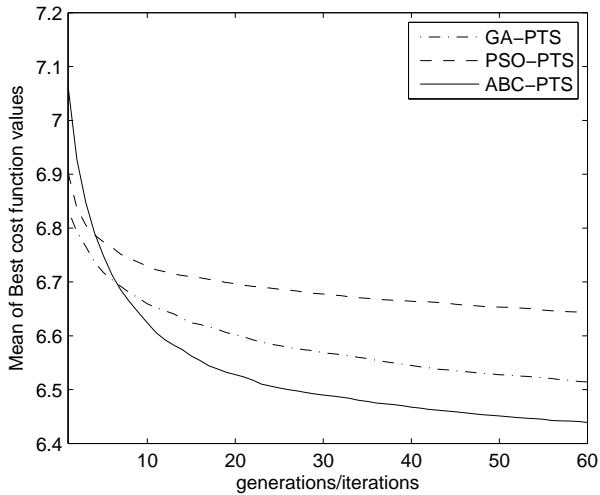}
\caption{Comparison of mean of best cost function values for  different swarm intelligence methods.}
\label{fig5}
\end{figure}
For three swarm intelligence algorithms, i.e. the  PSO-PTS~\cite{IEEEconf:20,IEEEconf:21},
the MDGA~\cite{IEEEconf:22,IEEEconf:23} and the ABC-PTS, 100 experiments are performed to compare  PAPR convergence performance
 for an OFDM symbol, where subblocks $M=16$ are generated by random partition, the same size of population is $S=P=30$ and the
same maximum iteration number $G=K=60$.
Fig.~\ref{fig5} shows the simulation results of three different methods on the mean of the best cost function values.
In initial phase (approximately $1-3$ iterations), the performance of the ABC-PTS is inferior to that of PSO-PTS and MDGA.
As the increase of iterations, the performance of the ABC-PTS is  better than that of PSO-PTS and MDGA.
 Although the PAPR performance is improved with the increase of
iteration number, the mean of PAPR getting by the iteration number $K=30$ is only less $0.1$\,dB than that of PAPR
getting by the iteration number $K=60$, so iteration number $K=30$ can be a suitable choice for our proposed ABC-PTS algorithm.

\section{Conclusion}
In this paper, we propose a  modified ABC based PTS algorithm
(ABC-PTS) to search better combination of phase factors for
OFDM signals. Compared to the existing PAPR reduction methods, the ABC-PTS algorithm can
get better PAPR reduction
and  significantly reduce the computational complexity for larger PTS
subblocks at the same time.
Moreover, because the ABC-PTS algorithm only has three control parameters, so it is easy to be
adjusted. Simulation results show that the ABC-PTS algorithm is an efficient
method which can provide a better PAPR performance.




\begin{thebibliography}{21}
\bibitem{IEEEconf:1}
Part 11: Wireless LAN Medium Acess Control (MAC) and Physical Layer
(PHY) Specifications: High-speed Physical Layer in the 5 GHz Band,
IEEE Standard 80.11a-1999.

\bibitem{IEEEconf:2}
Local and Metropolitan Area Networks-Part 16, Air Interface for Fixed Broadband
Wireless Access System, IEEE Standard 802.16a.

\bibitem{IEEEconf:3}
U. Reimers, `Digital Video Broadcasting,''
\emph{IEEE Commun. Mag.}, vol.~36, no.~6, pp.~104-110, June. 1998.

\bibitem{IEEEconf:4}
X.~Li, L. J. Cimini, Jr, ``Effect of clipping and filtering on the
performance of OFDM,'' \emph{IEEE Commun. Lett.}, vol.~2, no.~5,
pp.~131-133, May. 1998.

\bibitem{IEEEconf:5}
J.~A.~Davis, and J.~Jedwab, ``Peak to mean power control in OFDM,
Golay complementary sequences and Reed-Miller codes,'' {\em IEEE
Trans. Inform. Theory,} vol.~45, no.~7, pp.~2397-2417, Nov. 1999.

\bibitem{IEEEconf:6}
W. Chen, and C. Tellambura, "Identifying a class of multiple shift
complementary sequences in the second order cosets of the first
order Reed-Muller codes," \emph{IEEE International Conference on
Communications (ICC)}, pp.618-621, 2005.

\bibitem{IEEEconf:7}
R. W. B$\ddot{\mathrm{a}}$ml, R. F. H. Fisher and J. B. Huber,
``Reducing the Peak-to-Average Power Ratio of multicarrier
Modulation by Selected Mapping,''  \emph{Elect. Lett.}, vol.~32,
no.~22, pp.~2056-57, Oct. 1996.

\bibitem{IEEEconf:8}
J. Tellado and J. M. Cioffi, ``Further results on peak-to-average
ratio reduction,'' ANSI Document, T1E1.4 no. 98-252, Aug. 1998.

\bibitem{IEEEconf:9}
J. Tellado, ``Peak to average power reduction for multicarrier
Modulation,'' Ph.D. dissertation, Stanford Univ., 2000.

\bibitem{IEEEconf:10}
D. Jones, ``Peak power reduction in OFDM and DMT via active channel
modification,'' in \emph{Proc. 33rd Asilomar Conf. Signals, Systems
and Computers 1999},  pp.~1076-1079,

\bibitem{IEEEconf:11}
B. S. Krongold and D. L. Jones, ``PAR reduction in OFDM via active
constellation extension,'' in \emph{Proc. IEEE Int. Conf. Acoustics,
Speech, and Signal Processing 2003}, pp.~IV525-IV528,

\bibitem{IEEEconf:12}
M. Sharif and B. Hassibi, ``Existence of codes with constant PMEPR
and ralated design,''  \emph{IEEE Trans. Signal Processing.},
vol.~52, no.~10, pp.~2836-2846, Oct. 2004.

\bibitem{IEEEconf:13}
Y. J. Kou, W.S. Lu and A. Antoniou, `` A new peak-to-average power
-ratio reduction algorithms for OFDM systems via constellation
extension,'' \emph{IEEE Trans. Wireless Commun.}, vol.~6, no.~5,
pp.~1823-1832, May. 2007.

\bibitem{IEEEconf:14}
S. H. M$\ddot{\mathrm{u}}$ller and J. B. Huber, ``OFDM with reduce
peak-to-average power ratio by optimum combination of partial
transmit sequences,''  \emph{Elect. Lett.}, vol.~33, no.~5,
pp.~368-369, Feb. 1997.

\bibitem{IEEEconf:15}
L. J. Cimini, Jr. and N. R. Sollenberger, ``Peak-to-average power
ratio reduction of an OFDM signal using partial transmit
sequences,'' \emph{IEEE Commun. Lett.}, vol.~4, no.~3, pp.~86-88,
Mar. 2000.

\bibitem{IEEEconf:16}
S.  H. Han and J.  H. Lee. ``PAPR reduction of OFDM signals using a
reduced complexity PTS technique,'' \emph{IEEE Signal Processing
Lett.}, vol. 11, no. 11, pp. 887-890, Nov. 2004.

\bibitem{IEEEconf:17}
C. Tellambura,  ``Improved phase factor computation for the PAR
reduction of an OFDM signal using PTS,'' \emph{IEEE Commun. Lett.},
vol.~5, no.~4, pp.~135-137, Apr.  2001.


\bibitem{IEEEconf:18}
C. Tellambura,  ``Computation of the continuous-time PAR of an OFDM
signal with BPSK subcarriers,'' \emph{IEEE Commun. Lett.}, vol.~5,
no.~5, pp.~185-187, May. 2001.
\bibitem{IEEEconf:19}
Tao Jiang, Weidong Xiang, P. C. Richardson, Jinhua Guo, and Guangxi Zhu,
``PAPR Reduction of OFDM Signals Using Partial Transmit Sequences With
Low Computational Complexity,'' \emph{IEEE Trans. Broadcasting.}, vol.~53,
no.~3, pp.~719-724, Sep. 2007.
\bibitem{IEEEconf:20}
Jyh-Hong Wen, Shu-Hong Lee, Yung-Fa Huang and Ho-Lung Hung, ``A suboptimal PTS algorithm
based on paticle swarm optimization for PAPR reduction in OFDM systems,'' EURASIP Journal on
wireless communication and networking, vol. 2008, Article No.14.

\bibitem{IEEEconf:21}
Ho-Lung Hung, Yung-Fa Huang, Cheng-Ming Yeh and Tan-Hsu Tan, ``performance of paticle swarm optimization
techniques on PAPR reduction for  OFDM systems,'' \emph{IEEE International Conference on
systems, Man and Cybernetics}, 2008 (SMC 2008) 12-15 Oct.2008, pp.~2390-2395.

\bibitem{IEEEconf:22}
Yang Zhang, Qiang Ni, Hsiao-Hwa Chen and Yonghua Song, ``An intelligent genetic algorithm for
PAPR reduction in a multi-carrier CDMA wireless system,'' \emph{IEEE International Conference on
wireless communications and mobile computing conference,}, 2008. IWCMC ' 08. 6-8 Aug.2008, pp.~1052-1057.

\bibitem{IEEEconf:23}
Yang Zhang, Qiang Ni, Hsiao-Hwa Chen, ``A new partial transmit sequence scheme using  genetic algorithm
for peak-to-average power ratio reduction
in a multi-carrier code division multiple access wireless system,'' \emph{International Journal of Autonomous
 and Adaptive Communications Systems Issue}, vol. 2, Number 1/2009, pp.~40-57.

\bibitem{IEEEconf:24}
D. Karaboga, ``An idea based on honey bee swarm for numerical
optimization,'' Technical Report-TR 06,
  Erciyes university,
Engineering Faculty, Computer Engineering Department, 2005.

\bibitem{IEEEconf:25}
B. Basturk, D. Karaboga, ``An artificial bee colony (ABC) algorithm
for numeric function optimization,'' \emph{IEEE Swarm Intelligence
Symposium}, 2006.

\bibitem{IEEEconf:26}
D. Karaboga, B. Basturk, ``A powerful and efficient algorithm for
numeric function optimization:artificial bee colony (ABC)
algorithm,'' \emph{Journal of Global Optimization}, vol. 39,
pp.~459-471, 2007.

\end{thebibliography}
%

\section*{Acknowledgment}
We would like to thank the anonymous referees for their great
constructive comments to improve our works.




\end{document}